\documentclass[12pt]{article}
\pdfoutput=1

\usepackage{amsmath}
\usepackage{amsfonts}
\usepackage{amscd}
\usepackage{amsthm}
\usepackage{setspace}

\usepackage{graphicx}
\usepackage{authblk}
\usepackage{caption}
\usepackage{ytableau}
\usepackage{mathtools}

\usepackage[OT2,T1]{fontenc}
\DeclareSymbolFont{cyrletters}{OT2}{wncyr}{m}{n}
\DeclareMathSymbol{\Sha}{\mathalpha}{cyrletters}{"58}

\def\Z{\mathbb{Z}}

\def\P{\mathbb{P}}

\def\til{\tilde}

\begin{document}

\begin{titlepage}

\begin{flushright}
KEK-TH-2141
\end{flushright}

\vskip 1cm

\begin{center}

{\bf \Large A note on transition in discrete gauge groups in F-theory}

\vskip 1.2cm

Yusuke Kimura$^1$ 
\vskip 0.6cm
{\it $^1$KEK Theory Center, Institute of Particle and Nuclear Studies, KEK, \\ 1-1 Oho, Tsukuba, Ibaraki 305-0801, Japan}
\vskip 0.4cm
E-mail: kimurayu@post.kek.jp

\vskip 2cm
\abstract{We observe a new puzzling physical phenomenon in F-theory on the multisection geometry, wherein a model without a gauge group transitions to another model with a discrete $\Z_n$ gauge group via Higgsing. This phenomenon may suggest an unknown aspect of F-theory compactification on multisection geometry lacking a global section. A possible interpretation of this puzzling physical phenomenon is proposed in this note. We also propose a possible interpretation of another unnatural physical phenomenon observed for F-theory on four-section geometry, wherein a discrete $\Z_2$ gauge group transitions to a discrete $\Z_4$ gauge group via Higgsing as described in the previous literature.}  

\end{center}
\end{titlepage}

\tableofcontents
\section{Introduction}
In F-theory formulation \cite{Vaf, MV1, MV2}, theories are compactified on spaces that admit torus fibration. The modular parameter of an elliptic curve as a fiber of this fibration is identified with the axiodilaton in type IIB superstring; this enables axiodilaton to have $SL_2(\Z)$ monodromy in F-theory formulation. 
\par There are cases in which genus-one fibrations admit a global section, and those in which they do not have a global section. These two cases have different physical interpretations, as we now explain. Genus-one fibrations with a global section are frequently referred to as elliptic fibrations in the F-theory literature. Geometrically, that a genus-one fibration has a global section means that one can choose a point for each elliptic fiber over every point in the base space, and one can move this chosen point throughout the base space, yielding a smooth copy of the base space. Thus, to have a global section implies that one has a copy of the base space embedded in the total elliptic fibration. 
\par F-theory models compactified on elliptic fibrations with a global section have been studied in the literature, e.g., in \cite{MorrisonPark, MPW, BGK, BMPWsection, CKP, BGK1306, CGKP, CKP1307, CKPS, AL, EKY1410, LSW, CKPT, CGKPS, MP2, MPT1610, BMW2017, CL2017, BMW1706, KimuraMizoguchi, Kimura1802, LRW2018, MizTani2018, CMPV1811, TT2019, Kimura1903, EJ1905, LW1905}. The number of $U(1)$ factors is identified with the rank of the group generated by the set of global sections \cite{MV2}, which is known as the ``Mordell--Weil group,'' and the rank of this group is positive when there are two or more independent sections. Therefore, the F-theory model has a $U(1)$ factor when it is compactified on an elliptic fibration with two (or more) independent sections. 
\par F-theory models on genus-one fibrations lacking a global section have attracted interest recently, for reasons including the discrete gauge group \footnote{For recent studies on discrete gauge groups, see, for example, \cite{KNPRR, ACKO, BS, HSsums, CIM, BISU, ISU, BCMRU, BCMU, MRV, HS, BRU, KKLM, HS2, GPR}.} that arises \cite{MTsection} in this type of F-theory compactification. A genus-one fibration without a section still admits a multisection, which ``wraps around'' over the base multiple times. Multisections of ``degree'' $n$, or simply ``$n$-sections,'' are those wrapped around over the base $n$ times and therefore they intersect with the fiber $n$ times. In F-theory on a genus-one fibration with a multisection, the degree of the multisection corresponds to the degree of a discrete gauge group \footnote{The discrete part of the ``Tate--Shafarevich group,'' which is often denoted as $\Sha$, of the Jacobian of the compactification space yields a discrete gauge group arising in F-theory compactification \cite{MTsection}. Given a Calabi--Yau genus-one fibration $Y$, the ``Tate--Shafarevich group,'' $\Sha(J(Y))$, of the Jacobian $J(Y)$ can be considered. The discrete part of this Tate--Shafarevich group $\Sha(J(Y))$ is then identified with the discrete gauge group arising in F-theory on the genus-one fibration $Y$ \cite{BDHKMMS}.} forming in F-theory \cite{MTsection}; a discrete $\Z_n$ gauge group arises in F-theory on a genus-one fibration with an $n$-section. Recent studies of F-theory compactifications on genus-one fibrations without a section can be found, e.g., in \cite{BM, MTsection, AGGK, KMOPR, GGK, MPTW, MPTW2, BGKintfiber, CDKPP, LMTW, K, K2, ORS1604, KCY4, CGP, Kdisc, Kimura1801, AGGO1801, Kimura1806, TasilectWeigand, CLLO, TasilectCL, HT, Kimura1810, Kimura1902, Kimura1905} \footnote{F-theory on genus-one fibrations lacking a global section was investigated in \cite{BEFNQ, BDHKMMS}.}. 

\vspace{5mm}

\par The aim of this note was to study the transitions in discrete gauge groups forming in F-theory on the moduli of multisection geometry and, as a result, we point out that we observed a physical phenomenon that might be interesting. 
\par A physically puzzling transition from a discrete $\Z_2$ gauge group to a discrete $\Z_4$ gauge group was discussed in \cite{Kimura1905}. Our analysis in this study finds that this is not the only puzzling phenomenon that can be observed in F-theory models on the multisection geometry.
\par We point out a new physical phenomenon that appears unnatural in this note, by studying the structure of the multisection geometry. There are various ways a multisection splits into multisections of smaller degrees. By analyzing these structures, and by considering the physical interpretation of these structures, we observe the new puzzling physical phenomenon. 
\par To be clear, the analysis of splitting of a multisection enables us to observe a transition process wherein an F-theory model that has neither a discrete gauge group nor a $U(1)$ appears to ``break down'' into another F-theory model with a discrete gauge group via Higgsing. Because an F-theory model transitions to another model with a discrete gauge group via Higgsing in this process, the original model is supposed to possess a larger gauge symmetry than the discrete gauge group that the model has after the transition. However, the original F-theory model appears not to have any gauge group, simply because it does not have either a discrete gauge group or $U(1)$. This observation poses us a question: how should we interpret this transition process, and what is the gauge group that the original model possesses breaking down into a discrete gauge group of another model? 
\par The argument that we use leading to this observation does not depend on the degree of a multisection; the observation generally holds for an F-theory model on any $n$-section geometry for $n\ge 3$. Given a generic F-theory model with a discrete $\Z_n$ gauge group for any $n\ge 3$, one can find some F-theory model without $U(1)$ or a discrete gauge group that undergoes transition to that model via Higgsing. 

\vspace{5mm}

\par In addition, we propose a possible physical interpretation that can explain a puzzling phenomenon observed in \cite{Kimura1905} in section \ref{sec3.1}. This proposed interpretation of a physically unnatural phenomenon hints at a physical interpretation of a puzzling phenomenon wherein an F-theory model without a discrete gauge group or $U(1)$ ``breaks down'' into another model with a discrete $\Z_n$ gauge group via Higgsing, which we have just mentioned. We give a description of this puzzling phenomenon in detail in section \ref{sec2.1}. We also discuss a possible interpretation of this puzzling physical phenomenon in section \ref{sec3.2}. 
\par When we propose possible physical interpretations of the phenomena that we have discussed, we also find new open problems. We mention these problems at the end of this study.
\par Our proposal in section \ref{sec3.1} of a possible interpretation of the physically unnatural phenomenon observed in \cite{Kimura1905} is tested along the bisection geometries loci in the four-section geometry in \cite{Kimura1908}.
\par The motivation of this note was to point out a new physical phenomenon that appears unnatural. We proposed possible interpretations of this puzzling phenomenon.
\par At the level of geometry, the arguments we use in this note do not depend on the dimensionality of the space. However, when we consider four-dimensional (4D) F-theory models, the issue of flux \cite{BB, SVW, W96, GVW, DRS} \footnote{See, e.g., \cite{MSSN, CS, MSSN2, BCV, MS, KMW, GH, KMW2, IJMMP, KSN, CGK, BCV2, LMTW, SNW} for recent progress on F-theory compactifications with four-form flux.} arises, and this can make the analysis complicated. To this end, we focus on six-dimensional (6D) F-theory models when we discuss possible physical interpretations of the puzzling phenomena in this study. \cite{Nak, DG, G} discussed structures of genus-one fibrations of 3-folds.
\par Recent studies on building F-theory models have focused on local F-theory models \cite{DWmodel, BHV1, BHV2, DW}. To discuss the issues of gravity and the issues of early universe such as inflation, however, the global aspects of the geometry need to be studied. The geometries of compactification spaces are analyzed from the global perspective here.

\vspace{5mm}

\par This note is structured as follows. In section \ref{sec2.1}, we describe a physically puzzling phenomenon in F-theory on multisection geometry, in which an F-theory model without $U(1)$ or a discrete gauge group appears to ``break down'' into another F-theory model with a discrete gauge group via Higgsing. We also briefly review the physically unnatural phenomenon pointed out in \cite{Kimura1905} in section \ref{sec2.2}. 
\par In section \ref{sec3.1}, we propose a possible physical interpretation that can explain the unnatural phenomenon discussed in \cite{Kimura1905}. We propose a possible physical interpretation of the unnatural phenomenon that we observed in the trisection geometry in section \ref{sec3.a}. Making use of these interpretations as a hint, we also propose a physical interpretation of the puzzling phenomenon that we describe in section \ref{sec2.1} in section \ref{sec3.2}. 
\par We discuss some questions that these interpretations raise and we also discuss open problems in section \ref{sec4}.

\section{Transitions in discrete gauge groups in multisection geometry}
\label{sec2}

\subsection{Puzzling transitions of models with discrete gauge groups}
\label{sec2.1}
We analyze the splitting process of multisections in the moduli of multisection geometry to observe a physical phenomenon that seems puzzling. As noted in the introduction, this is new and different from the unnatural physical phenomenon pointed out in \cite{Kimura1905}. 
\par We would like to demonstrate that when we study a certain splitting of an $n$-section, namely a multisection that intersects with a fiber $n$ times, the analysis leads to a physically puzzling phenomenon. 
\par A generic member of the $n$-section geometry does not have a global section, and it only contains an $n$-section as a multisection of the smallest degree. However, when the coefficients of the equation of the genus-one fibration as parameters take special values, so certain polynomials, in the variables of the function field of the base space, plugged into the coordinate variables satisfy the equation of the genus-one fibration, this situation geometrically implies that the genus-one fibration admits a global section \cite{K}. This is because this ``solution'' specifies a point in each fiber over the base space, yielding a copy of the base. 
\par At this special point in the moduli of $n$-section geometry where the equation of the genus-one fibration admits a ``solution'' yielding a global section, an $n$-section splits into a global section and an $(n-1)$-section. For the case $n=2$, a bisection simply splits into a pair of global sections. For $n\ge 3$, this process of the splitting of an $n$-section leads to an interpretation that appears physically puzzling, as we shall shortly show.
\par A discrete $\Z_n$ gauge group in F-theory forms on a genus-one fibration with an $n$-section \cite{MTsection} as mentioned in the introduction. When this $n$-section splits into a global section and an $(n-1)$-section in the moduli, the Mordell--Weil rank of the resulting elliptic fibration is zero. (This is because it only has one independent global section, i.e., the ``zero-section.'') For this reason, F-theory on the resulting elliptic fibration has neither a discrete gauge group nor $U(1)$. 
\par We would like to demonstrate these statements for $n=3, 4$, namely for the trisection geometry and the four-section geometry. Trisection geometry built as cubic hypersurfaces in $\P^2$ fibered over a base space was considered in \cite{KMOPR}. We consider special points in the complex structure moduli of the trisection geometry where the trisection splits into a global section and a bisection. Specifically, we consider the cubic hypersurfaces fibered over a base space given by the following equation:
\begin{equation}
\label{cubic hypersurface in 2.1}
c_1\, X^3+c_2\, X^2Y+ c_3\, XY^2+c_4\, Z^3+ c_5\, Z^2X+ c_6\, ZX^2+ c_7\, Z^2Y+ c_8\, ZY^2 + c_9\, XYZ=0.
\end{equation}
$[X:Y:Z]$ denotes homogeneous coordinates on $\P^2$, and $c_i$, $i=1, \ldots, 9$, are sections of line bundles of the base space. (Sections of line bundles $c_i$, $i=1, \ldots, 9$, are subject to the conditions such that the total space of fibration is Calabi--Yau \cite{KMOPR}.) From this equation, one can find that the Calabi--Yau genus-one fibration (\ref{cubic hypersurface in 2.1}) has a constant section $[X:Y:Z]=[0:1:0]$. Trisection geometry built as cubic hypersurface in $\P^2$ fibered over a base space generically contains a trisection and it lacks a global section \cite{KMOPR}, and the trisection splits into a global section and a bisection at the special point (\ref{cubic hypersurface in 2.1}) in the complex structure moduli. The Mordell--Weil group of the Calabi--Yau hypersurface (\ref{cubic hypersurface in 2.1}) consists only of the constant zero-section $[0:1:0]$; therefore, the Mordell--Weil group is 0. 
\par The Jacobian fibration \footnote{Construction of the Jacobian fibration of an elliptic curve can be found in \cite{Cas}.} of the trisection geometry built as a cubic hypersurface in $\P^2$ fibered over a base space is given in \cite{KMOPR}. By plugging the relation, that the coefficient of the term $Z^3$ vanishes, into the Weierstrass coefficients given in \cite{KMOPR}, one obtains the Weierstrass coefficients of the elliptic fibration (\ref{cubic hypersurface in 2.1}). The Weierstrass coefficients $f,g$ of $y^2=x^3+fx+g$ of the elliptic fibration (\ref{cubic hypersurface in 2.1}) are given as
\begin{eqnarray}
\label{Weierstrass cubic hypersurface in 2.1}
f= & 3\cdot \Big[ -\big(c_9^2-4(c_6c_8+c_3c_5+c_2c_7)\big)^2 \\ \nonumber
& +24[-c_9 (c_4c_2c_3+c_2c_8c_5+c_3c_6c_7+c_1c_8c_7) \\ \nonumber
& +2\big(c_3^2c_6c_4+c_1c_8^2c_5+c_2c_3c_5c_7+c_1c_3c_7^2 +c_8(c_2^2c_4-3c_1c_3c_4+c_3c_6c_5+c_2c_6c_7)\big)] \Big]. \\ \nonumber
g= & 2\cdot \Big[ \big(c_9^2-4(c_6c_8+c_3c_5+c_2c_7)\big)^3 \\ \nonumber
& -36\big(c_9^2-4(c_6c_8+c_3c_5+c_2c_7)\big)\Big(-c_9(c_2c_3c_4+c_2c_8c_5+c_3c_6c_7+c_1c_8c_7) \\ \nonumber
& +2\big(c_3^2c_6c_4+c_1c_8^2c_5+c_2c_3c_5c_7+c_1c_3c_7^2 \\ \nonumber
& +c_8(c_2^2c_4-3c_1c_3c_4+c_3c_6c_5+c_2c_6c_7)\big) \Big) \\ \nonumber
& +216\Big( (c_2c_3c_4+c_2c_8c_5+c_3c_6c_7+c_1c_8c_7)^2 \\ \nonumber
& +4\big( -c_1c_3^3c_4^2-c_1^2c_8^3c_4+c_3^2c_4(-c_2c_6+c_1c_9)c_7-c_1c_3^2c_5c_7^2 \\ \nonumber
& -c_8^2(c_4(c_2^2c_6-2c_1c_3c_6-c_1c_2c_9)+c_1c_5(c_3c_5+c_2c_7)) \\ \nonumber
& -c_3c_8(c_4(-c_2c_6c_9+c_1c_9^2+c_2^2c_5+c_3(c_6^2-2c_1c_5)+c_1c_2c_7)+c_7(c_2c_6c_5-c_1c_9c_5+c_1c_6c_7)) \big) \Big) \Big].
\end{eqnarray}
\par The discriminant of the Weierstrass equation whose coefficients are given as in (\ref{Weierstrass cubic hypersurface in 2.1}) is irreducible, consisting of a single factor. It follows that the Jacobian (\ref{Weierstrass cubic hypersurface in 2.1}) has no singularities, canonical or terminal, for generic values of the parameters $c_1, \ldots, c_9$. Therefore, the elliptic fibration (\ref{cubic hypersurface in 2.1}) has no gauge symmetry in F-theory. This can be demonstrated as follows: One can consider a special situation where $c_2, c_5, c_6, c_7$ vanish. For this situation, the Weierstrass equation of the Jacobian reduces to the following form:
\begin{equation}
y^2= x^3+ 3(-c_9^4-144c_1c_3c_4c_8)\, x+ 2\big(c_9^6-72c_9^2(-3c_1c_3c_4c_8)+864(-c_1c_3^3c_4^2-c_1^2c_8^3c_4-c_1c_3c_4c_8c_9^2)\big).
\end{equation}
The discriminant of this reduced Weierstrass equation is given as follows:
\begin{equation}
\label{discriminant c2567 in 2.1}
\begin{split}
\Delta(c_2=c_5=c_6=c_7=0) \sim & \, c_1 c_4 (-c_3 c_8 c_9^8 - c_1 c_8^3 c_9^6 - c_3^3 c_4 c_9^6 + 207 c_1 c_3^2 c_4 c_8^2 c_9^4 \\
& + 648 c_1^2 c_3 c_4 c_8^4 c_9^2 + 648 c_1 c_3^4 c_4^2 c_8 c_9^2 + 432 c_1^3 c_4 c_8^6 \\
& + 432 c_1 c_3^6 c_4^3 - 864 c_1^2 c_3^3 c_4^2 c_8^3).
\end{split}
\end{equation}
The right-hand side of (\ref{discriminant c2567 in 2.1}) yields factorization into irreducible factors, and from the discriminant (\ref{discriminant c2567 in 2.1}) one can learn that there are three possibilities for the original discriminant of the Jacobian (\ref{Weierstrass cubic hypersurface in 2.1}): (i) the discriminant factors into two linear factors that reduce to $c_1$ and $c_4$, respectively, under $c_2=c_5=c_6=c_7=0$, and an irreducible factor; (ii) the discriminant factors into a degree-two factor that reduces to $c_1c_4$ under $c_2=c_5=c_6=c_7=0$, and an irreducible factor; (iii) the discriminant is irreducible (and it does not vanish to order greater than one). 
\par One can exclude the first two possibilities, (i) and (ii), by considering another situation where some of the coefficients $c_i$ vanish. If (i) occurs, the discriminant of the Jacobian (\ref{Weierstrass cubic hypersurface in 2.1}) contains linear factors of the forms: $c_1+\alpha_1 c_2+\alpha_2 c_5 +\alpha_3 c_6+ \alpha_4 c_7$ and $c_4+\alpha_5 c_2+\alpha_6 c_5 +\alpha_7 c_6+ \alpha_8 c_7$, where $\alpha_1, \ldots, \alpha_8$ are constants. Then, we consider the following situation: $c_2=c_5=c_6=c_8=0$. Under this situation, linear factors reduce as follows: $c_1+\alpha_4 c_7$ and $c_4+\alpha_8 c_7$.
\par Under the situation $c_2=c_5=c_6=c_8=0$, the Jacobian (\ref{Weierstrass cubic hypersurface in 2.1}) reduces to the following Weierstrass equation:
\begin{equation}
y^2= x^3+ 3(-c_9^4+48c_1c_3c_7^2)\, x+ 2\big(c_9^6-72c_9^2c_1c_3c_7^2+864(-c_1c_3^3c_4^2+c_3^2c_1c_4c_7c_9)\big).
\end{equation}
The discriminant for this situation is given as follows:
\begin{equation}
\label{discriminant c2568 in 2.1}
\begin{split}
\Delta(c_2=c_5=c_6=c_8=0) \sim & \,  c_1 c_3^2 (c_4 c_7 c_9^7 - c_3 c_4^2 c_9^6 - c_1 c_7^4 c_9^4 - 72 c_1 c_3 c_4 c_7^3 c_9^3 + 504 c_1 c_3^2 c_4^2 c_7^2 c_9^2 \\
& - 864 c_1 c_3^3 c_4^3 c_7 c_9 + 64 c_1^2 c_3 c_7^6 + 432 c_1 c_3^4 c_4^4).
\end{split}
\end{equation}
The right-hand side of (\ref{discriminant c2568 in 2.1}) yields factorization into irreducible factors. However, the discriminant (\ref{discriminant c2568 in 2.1}) does not contain factor $c_4+\alpha_8 c_7$, which is a contradiction. Thus, (i) does not occur. (ii) can be excluded in a similar fashion, and we find that (iii) is a unique possibility. Therefore, we can conclude that the discriminant of the Jacobian (\ref{Weierstrass cubic hypersurface in 2.1}) is irreducible, and there is not any codimension-one locus in the base space at which the discriminant vanishes to order greater than one.
 
\par For the four-section geometry, by applying an argument similar to that we have just given for the trisection geometry, the fibrations in the complex structure moduli of the four-section geometry where the four-section splits into a global section and a trisection have no gauge group in F-theory. The genus-one curve obtained as a complete intersection of two quadric hypersurfaces in $\P^3$, fibered over any base space, yields a genus-one fibration with a four-section \cite{BGKintfiber, Kdisc, Kimura1905}. This genus-one fibration with a four-section is described by the following equation in a general form:
\begin{eqnarray}
\label{complete intersection general in 2.2}
a_1\, x_1^2+a_2 \,  x_2^2 + a_3 \,  x_3^2+a_4 \, x_4^2 + 2a_5\, x_1x_2+2a_6\, x_1x_3 + & \\ \nonumber
2a_7 \, x_1x_4+2a_8 \, x_2x_3 +2a_9\, x_2x_4 +2a_{10}\, x_3x_4 & =0 \\ \nonumber
b_1\, x_1^2+b_2\,  x_2^2 + b_3\,  x_3^2+b_4 \, x_4^2 + 2b_5\, x_1x_2+2b_6\, x_1x_3 + & \\ \nonumber
2b_7\, x_1x_4+2b_8\, x_2x_3 +2b_9\, x_2x_4 +2b_{10}\, x_3x_4 & =0.
\end{eqnarray}
$[x_1:x_2:x_3:x_4]$ denotes the coordinates of $\P^3$, and $a_i, b_j$, $i,j=1, \ldots, 10$, are sections of line bundles of the base space \footnote{Certain conditions are imposed on the sections of line bundles, so when the Jacobian fibration is taken, the Weierstrass form of which is given by $y^2=x^3+fx+g$, then the associated divisors $[f], [g]$, satisfy $[f]=-4K$, $[g]=-6K$, where $K$ denotes the canonical divisor of the base, to ensure that the total space of fibration is Calabi--Yau, as described in \cite{Kimura1905}.}. The Jacobian fibration of this genus-one fibration (\ref{complete intersection general in 2.2}) always exists, the types of the singular fibers and the discriminant locus of which are identical to the genus-one fibration. See, e.g., \cite{BM, Kimura1905} for the construction of the Jacobian fibration of the genus-one fibration (\ref{complete intersection general in 2.2}).
\par We particularly consider the special locus in the four-section geometry (\ref{complete intersection general in 2.2}) where the coefficients satisfy the following relations: 
\begin{eqnarray}
\label{global section locus in complete intersection in 2.1}
a_1=a_2 =a_5 \\ \nonumber
b_1=b_2 =b_5.
\end{eqnarray}
Along the locus (\ref{global section locus in complete intersection in 2.1}), the complete intersection (\ref{complete intersection general in 2.2}) becomes 
\begin{eqnarray}
\label{complete intersection with a global section in 2.1}
a_1\, (x_1+x_2)^2+ a_3 \,  x_3^2+a_4 \, x_4^2 +2a_6\, x_1x_3 + & \\ \nonumber
2a_7 \, x_1x_4+2a_8 \, x_2x_3 +2a_9\, x_2x_4 +2a_{10}\, x_3x_4 & =0 \\ \nonumber
b_1\, (x_1+x_2)^2+ b_3\,  x_3^2+b_4 \, x_4^2 +2b_6\, x_1x_3 + & \\ \nonumber
2b_7\, x_1x_4+2b_8\, x_2x_3 +2b_9\, x_2x_4 +2b_{10}\, x_3x_4 & =0.
\end{eqnarray}
Thus, the complete intersections belonging to the locus (\ref{global section locus in complete intersection in 2.1}) admit a constant section $[x_1:x_2:x_3:x_4]=[1:-1:0:0]$. The four-section splits into a global section and a trisection along the locus (\ref{global section locus in complete intersection in 2.1}). The Mordell--Weil group of the complete intersection (\ref{complete intersection with a global section in 2.1}) consists only of the zero-section $[1:-1:0:0]$, and therefore it is 0. 
\par Using the techniques discussed in \cite{BM, Kimura1905}, one can construct the double cover of the quartic polynomial from the complete intersection (\ref{complete intersection with a global section in 2.1}), by subtracting the second equation times $\lambda$ from the first equation in (\ref{complete intersection with a global section in 2.1}) and arranging the coefficients of the resulting equation into a symmetric $4 \times 4$ matrix, then taking the double cover of the determinant of the matrix. The Jacobian fibration \cite{BM, MTsection} of the resulting double cover gives \cite{BM} the Jacobian fibration of the complete intersection (\ref{complete intersection with a global section in 2.1}). After computation, we find that the coefficients of the quartic polynomial of the double cover $\tau^2=e_0\lambda^4+ e_1\lambda^3+ e_2\lambda^2+ e_3\lambda+ e_4$ are 
\begin{eqnarray}
e_0= & (b_7b_8-b_6b_9)^2-b_1b_4(b_6-b_8)^2-b_1b_3(b_7-b_9)^2+2b_1b_{10}(b_6b_7-b_7b_8-b_6b_9+b_8b_9) \\ \nonumber
e_1= & -2 a_7 b_7 b_8^2 - 2 a_6 b_6 b_9^2 - 2 a_8 b_7^2 b_8 + 2 a_9 b_6 b_7 b_8 - 2 a_9 b_6^2 b_9 + 2 a_8 b_6 b_7 b_9 \\ \nonumber
& + 2 a_7 b_6 b_8 b_9 + 2 a_6 b_7 b_8 b_9+ a_4 b_1 b_6^2 + a_1 b_4 b_6^2 + a_3 b_1 b_7^2 + a_1 b_3 b_7^2 \\ \nonumber
& + a_4 b_1 b_8^2 + a_1 b_4 b_8^2 + a_3 b_1 b_9^2 + a_1 b_3 b_9^2 - 2 b_{10} a_7 b_1 b_6 + 2 b_{10} a_9 b_1 b_6 \\ \nonumber
& + 2 a_6 b_1 b_4 b_6 - 2 a_8 b_1 b_4 b_6 - 2 b_{10} a_6 b_1 b_7 + 2 b_{10} a_8 b_1 b_7 + 2 a_7 b_1 b_3 b_7 \\ \nonumber
& - 2 a_9 b_1 b_3 b_7 - 2 b_{10} a_1 b_6 b_7 - 2 a_{10} b_1 b_6 b_7 + 2 b_{10} a_7 b_1 b_8 - 2 b_{10} a_9 b_1 b_8 \\ \nonumber
& - 2 a_6 b_1 b_4 b_8 + 2 a_8 b_1 b_4 b_8 - 2 a_4 b_1 b_6 b_8 - 2 a_1 b_4 b_6 b_8 + 2 b_{10} a_1 b_7 b_8 + 2 a_{10} b_1 b_7 b_8 \\ \nonumber
& + 2 b_{10} a_6 b_1 b_9 - 2 b_{10} a_8 b_1 b_9 - 2 a_7 b_1 b_3 b_9 + 2 a_9 b_1 b_3 b_9 + 2 b_{10} a_1 b_6 b_9 + 2 a_{10} b_1 b_6 b_9 \\ \nonumber
& - 2 a_3 b_1 b_7 b_9 - 2 a_1 b_3 b_7 b_9 - 2 b_{10} a_1 b_8 b_9 - 2 a_{10} b_1 b_8 b_9 \\ \nonumber
e_2= & a_9^2 b_6^2 + a_8^2 b_7^2 + a_7^2 b_8^2 + a_6^2 b_9^2 - 2 a_8 a_9 b_6 b_7 - 2 a_7 a_9 b_6 b_8 \\ \nonumber
& + 4 a_7 a_8 b_7 b_8 - 2 a_6 a_9 b_7 b_8 - 2 a_7 a_8 b_6 b_9 + 4 a_6 a_9 b_6 b_9 - 2 a_6 a_8 b_7 b_9 - 2 a_6 a_7 b_8 b_9 \\ \nonumber
& -a_1 a_4 b_6^2 - a_1 a_3 b_7^2 - a_1 a_4 b_8^2 - a_1 a_3 b_9^2 + 2 b_{10} a_6 a_7 b_1 - 2 b_{10} a_7 a_8 b_1 \\ \nonumber
& - 2 b_{10} a_6 a_9 b_1 + 2 b_{10} a_8 a_9 b_1 - a_7^2 b_1 b_3 - a_9^2 b_1 b_3 + 2 a_7 a_9 b_1 b_3 - a_6^2 b_1 b_4 - a_8^2 b_1 b_4 \\ \nonumber
& + 2 a_6 a_8 b_1 b_4 + 2 b_{10} a_1 a_7 b_6 - 2 b_{10} a_1 a_9 b_6 - 2 a_4 a_6 b_1 b_6 + 2 a_{10} a_7 b_1 b_6 + 2 a_4 a_8 b_1 b_6 \\ \nonumber
& - 2 a_{10} a_9 b_1 b_6 - 2 a_1 a_6 b_4 b_6 + 2 a_1 a_8 b_4 b_6 + 2 b_{10} a_1 a_6 b_7 - 2 b_{10} a_1 a_8 b_7 \\ \nonumber
& + 2 a_{10} a_6 b_1 b_7 - 2 a_3 a_7 b_1 b_7 - 2 a_{10} a_8 b_1 b_7 + 2 a_3 a_9 b_1 b_7 - 2 a_1 a_7 b_3 b_7 + 2 a_1 a_9 b_3 b_7 \\ \nonumber
& + 2 a_{10} a_1 b_6 b_7 - 2 b_{10} a_1 a_7 b_8 + 2 b_{10} a_1 a_9 b_8 + 2 a_4 a_6 b_1 b_8 - 2 a_{10} a_7 b_1 b_8 - 2 a_4 a_8 b_1 b_8 \\ \nonumber
& + 2 a_{10} a_9 b_1 b_8 + 2 a_1 a_6 b_4 b_8 - 2 a_1 a_8 b_4 b_8 + 2 a_1 a_4 b_6 b_8 - 2 a_{10} a_1 b_7 b_8 - 2 b_{10} a_1 a_6 b_9 + 2 b_{10} a_1 a_8 b_9 \\ \nonumber
& - 2 a_{10} a_6 b_1 b_9 + 2 a_3 a_7 b_1 b_9 + 2 a_{10} a_8 b_1 b_9 - 2 a_3 a_9 b_1 b_9 + 2 a_1 a_7 b_3 b_9 - 2 a_1 a_9 b_3 b_9 \\ \nonumber
& - 2 a_{10} a_1 b_6 b_9 + 2 a_1 a_3 b_7 b_9 + 2 a_{10} a_1 b_8 b_9 \\ \nonumber
e_3= & - 2 a_6 a_9^2 b_6  + 2 a_7 a_8 a_9 b_6  - 2 a_7 a_8^2 b_7  + 2 a_6 a_8 a_9 b_7  - 2 a_7^2 a_8 b_8  + 2 a_6 a_7 a_9 b_8  \\ \nonumber
& + 2 a_6 a_7 a_8 b_9  - 2 a_6^2 a_9 b_9 - 2 b_{10} a_1 a_6 a_7 + 2 b_{10} a_1 a_7 a_8 + 2 b_{10} a_1 a_6 a_9 - 2 b_{10} a_1 a_8 a_9 \\ \nonumber
& + a_4 a_6^2 b_1 + a_3 a_7^2 b_1 + a_4 a_8^2 b_1 + a_3 a_9^2 b_1 - 2 a_{10} a_6 a_7 b_1 - 2 a_4 a_6 a_8 b_1 \\ \nonumber
& + 2 a_{10} a_7 a_8 b_1 + 2 a_{10} a_6 a_9 b_1 - 2 a_3 a_7 a_9 b_1 - 2 a_{10} a_8 a_9 b_1 + a_1 a_7^2 b_3 + a_1 a_9^2 b_3 \\ \nonumber
& - 2 a_1 a_7 a_9 b_3 + a_1 a_6^2 b_4 + a_1 a_8^2 b_4 - 2 a_1 a_6 a_8 b_4 + 2 a_1 a_4 a_6 b_6 - 2 a_{10} a_1 a_7 b_6 \\ \nonumber
& - 2 a_1 a_4 a_8 b_6 + 2 a_{10} a_1 a_9 b_6 - 2 a_{10} a_1 a_6 b_7 + 2 a_1 a_3 a_7 b_7 + 2 a_{10} a_1 a_8 b_7 - 2 a_1 a_3 a_9 b_7 \\ \nonumber
& - 2 a_1 a_4 a_6 b_8 + 2 a_{10} a_1 a_7 b_8 + 2 a_1 a_4 a_8 b_8 - 2 a_{10} a_1 a_9 b_8 + 2 a_{10} a_1 a_6 b_9 - 2 a_1 a_3 a_7 b_9 \\ \nonumber
& - 2 a_{10} a_1 a_8 b_9 + 2 a_1 a_3 a_9 b_9 \\ \nonumber
e_4= & (a_7a_8-a_6a_9)^2-a_1a_4(a_6-a_8)^2-a_1a_3(a_7-a_9)^2+2a_1a_{10}(a_6a_7-a_7a_8-a_6a_9+a_8a_9)
\end{eqnarray}
By computing the discriminant and the Weierstrass coefficients of the Jacobian fibration of the resulting double cover, we deduce that the complete intersection (\ref{complete intersection with a global section in 2.1}) has no singularities, canonical or terminal. Therefore, we find that the fibrations (\ref{complete intersection with a global section in 2.1}) along the locus (\ref{global section locus in complete intersection in 2.1}) have no gauge symmetry in F-theory.

\par When we view the process of moving from a genus-one fibration with an $n$-section to an elliptic fibration with a global section and an $(n-1)$-section via the splitting of an $n$-section in the moduli from a physical viewpoint, which reverses the geometric process, an F-theory model without $U(1)$ or a discrete gauge group transitions to another F-theory model with a discrete $\Z_n$ gauge group via Higgsing, along the lines of the argument as in \cite{MTsection}. 
\par This physical interpretation appears puzzling: the original F-theory model is supposed to possess some gauge group, which breaks down into a discrete $\Z_n$ gauge group via Higgsing. However, the original F-theory model has neither $U(1)$ nor a discrete gauge symmetry, and we do not see any gauge group that can break down into a discrete $\Z_n$ gauge group. Although this must be a Higgsing process, it appears a discrete gauge group ``emerged from nothing.'' 
\par We propose a possible interpretation in section \ref{sec3.2} that might explain this puzzling phenomenon. We also discuss possible explanations for the unnatural phenomenon observed in \cite{Kimura1905} in section \ref{sec3.1}. An interpretation of the unnatural phenomenon observed in \cite{Kimura1905} that we describe in section \ref{sec3.1} gives a hint to interpret the puzzling phenomenon that we have just discussed. We describe a possible interpretation of the puzzle for the trisection model, which corresponds to the situation $n=3$, in section \ref{sec3.a}.
\par The possible interpretations that we propose in section \ref{sec3} also yield new questions and open problems. We mention this in section \ref{sec4}.

\vspace{5mm}

\par Before we move on to the next section, we would like to make a remark: if the degree $p$ of a discrete gauge group $\Z_p$ is prime (forming in F-theory on a genus-one fibration with a $p$-section), when the $p$-section splits into multisections of smaller degrees (which are not necessarily global sections), say $m$- and $l$-sections, because $p$ is prime and $m+l=p$, $m$ and $l$ should be coprime. Thus, they in fact generate a global section: whenever the $p$-section of prime degree splits into multisections of smaller degrees, they generate a global section. (The specific case $p=5$ is discussed in \cite{Kimura1905}. The statement is trivial for $p=2,3$.) For these situations (when $p\ne 2$), F-theory on the resulting elliptic fibration generically does not have a $U(1)$ factor because one expects only one independent section for the resulting elliptic fibration and, thus, the Mordell--Weil rank of the elliptic fibration is zero. The physical viewpoint of a $p$-section of prime degree splitting into multisections of smaller degrees yields a Higgsing process wherein an F-theory model without $U(1)$ or a discrete gauge group transitions to another F-theory model with a discrete $\Z_p$ gauge group. 

\subsection{Review of transitions in discrete $\Z_4$ gauge group}
\label{sec2.2}
We review the physically unnatural process discussed in \cite{Kimura1905} in which an F-theory model with a discrete $\Z_2$ gauge group transitions to another F-theory model with a discrete $\Z_4$ gauge group. 
\par A discrete $\Z_4$ gauge group forms in F-theory on a genus-one fibration with a four-section. For generic values of the parameters $a_i, b_j$, the genus-one fibration (\ref{complete intersection general in 2.2}) has a four-section to the fibration, but when the parameters take special values, the four-section splits into a pair of bisections as observed in \cite{Kdisc, Kimura1905}. Every bisection geometries locus in the four-section geometry (\ref{complete intersection general in 2.2}) admits a transformation to the following locus \cite{Kimura1908} :
\begin{equation}
\label{condition splitting in 2.2}
a_1=a_2=a_4=a_7=a_9=0.
\end{equation}
(The parameters, $a_3, a_5, a_6, a_8, a_{10}$, $b_j$, $j=1, \ldots, 10$, remain free.) 
\par The process of the four-section splitting into two bisections can be viewed from a physical viewpoint, which reverses the geometric order along the lines of the argument as in \cite{MTsection}, and a discrete $\Z_2$ gauge group rather appears ``enhanced'' to a discrete $\Z_4$ gauge group via Higgsing \cite{Kimura1905}. 
\par We propose a possible interpretation of this phenomenon in section \ref{sec3.1}. An interpretation that we propose in section \ref{sec3.1} can be used as a hint to interpret the puzzling phenomenon that we previously discussed in section \ref{sec2.1}. Utilizing interpretations that we give in section \ref{sec3.1} and in section \ref{sec3.a} as a hint, we propose a possible interpretation of the phenomenon discussed in section \ref{sec2.1} in section \ref{sec3.2}.

\section{Physical interpretations of the puzzles}
\label{sec3}

\subsection{An interpretation of the transition from a discrete $\Z_2$ gauge group to a discrete $\Z_4$ gauge group}
\label{sec3.1}
We propose a physical interpretation of the puzzle pointed out in \cite{Kimura1905}, namely a physically unnatural phenomenon wherein a discrete $\Z_2$ gauge group appears to be ``enhanced'' to a discrete $\Z_4$ gauge group via Higgsing.
\par First, we consider in four-section geometry a specific form of complete intersection of two quadric hypersurfaces. The consideration of this example provides a motivation for our interpretation of the puzzle in \cite{Kimura1905}, which we discuss shortly after we consider the example. We take the specific example as a starting point. We consider the complete intersection of the two quadric hypersurfaces given by the following equations:
\begin{eqnarray}
\label{specific complete intersection in 3.1}
x_1^2 + x_3^2 + 2\, f\, x_2x_4 = & 0 \\ \nonumber
x_2^2 + x_4^2 + 2\, g\, x_1x_3 = & 0.
\end{eqnarray}
As noted in section \ref{sec2.1}, $[x_1:x_2:x_3:x_4]$ denotes the homogeneous coordinates of $\P^3$. Here $f,g$ are sections of line bundles over the base space. This complete intersection can be defined over any base space. The complete intersections of this form (\ref{specific complete intersection in 3.1}) were considered in \cite{K2, Kdisc, Kimura1905}. The form (\ref{specific complete intersection in 3.1}) corresponds to a special case of the more general form (\ref{complete intersection general in 2.2}) that we discussed previously. A four-section splits into a pair of bisections in the complete intersections (\ref{specific complete intersection in 3.1}). Bisections are given by \cite{Kdisc} $\{x_1=0, \hspace{2mm} x_2=ix_4\}$ and $\{x_1=0, \hspace{2mm} x_2=-ix_4\}$.
\par The Jacobian fibration of the complete intersection (\ref{specific complete intersection in 3.1}) is given by \cite{K2, Kdisc}
\begin{equation}
\label{Jacobian of specific form in 3.1}
\tau^2=g^2\, \lambda^4 - (f^2g^2+1)\, \lambda^2 + f^2.
\end{equation}
(By a computation similar to those in \cite{K2, Kdisc}, we subtract the second equation times $\lambda$ as a variable from the first equation, then we arrange the coefficients of the resulting equation into a symmetric $4 \times 4$ matrix, as 
$$\begin{pmatrix}
1 & 0& -\lambda\, g & 0 \\
0 & -\lambda & 0 & f \\
-\lambda\, g & 0 & 1 & 0 \\
0 & f & 0 & -\lambda 
\end{pmatrix}. $$
Taking the double cover of the determinant of this matrix, we arrive at the equation (\ref{Jacobian of specific form in 3.1}). We absorbed the minus sign on the right-hand side by replacing $\tau$ with $i\, \tau$.)
\par This Jacobian fibration can be transformed into the (general) Weierstrass form as \cite{Kdisc}
\begin{equation}
\label{general Weierstrass form in 3.1}
y^2 = \frac{1}{4} x^3 - \frac{1}{2} (f^2g^2+1)\, x^2 + \frac{1}{4} (f^2g^2-1)^2\, x.
\end{equation}
The discriminant of the complete intersection (\ref{specific complete intersection in 3.1}) and the Jacobian fibration (\ref{Jacobian of specific form in 3.1}) is given by \cite{Kdisc}
\begin{equation}
\label{discriminant in 3.1}
\Delta = 16\, f^2g^2\, (f^2g^2-1)^4 = 16\, f^2g^2\, (fg-1)^4 \, (fg+1)^4.
\end{equation}
The 7-branes are wrapped on the components of the vanishing locus of the discriminant (\ref{discriminant in 3.1}). The fibers lying over the seven-branes wrapped on the components $\{f=0\}$ and $\{g=0\}$ are type $I_2$, and the fibers lying over the seven-branes wrapped on the components $\{fg-1=0\}$ and $\{fg+1=0\}$ have type $I_4$. From the general Weierstrass form, it can be confirmed \cite{Kdisc} that the type $I_4$ fibers over the seven-branes wrapped on the components $\{fg-1=0\}$ and $\{fg+1=0\}$ are split \cite{BIKMSV}.
\par At the geometric level, the arguments of this example apply to both 4D and 6D F-theory compactifications. If we consider the physics of the transition of discrete gauge groups in 4D F-theory, however, the structure of the flux also needs to be studied. To this end, we only consider 6D F-theory models here as noted in the introduction. 
\par As we explained previously, the complete intersection (\ref{specific complete intersection in 3.1}) is a bisection geometry. The condition where a bisection splits into a pair of global sections for bisection geometry is given in \cite{MTsection}. As described in \cite{BM, MTsection}, a genus-one fibration with a bisection is given by the double cover of a quartic polynomial:
\begin{equation}
\label{double cover in 3.1}
\tau^2 = e_0\, \lambda^4+e_1\, \lambda^3+e_2\, \lambda^2+e_3\, \lambda+e_4,
\end{equation}
and when the parameter $e_4$ in the double cover (\ref{double cover in 3.1}) becomes a perfect square, bisection of the genus-one fibration splits into two global sections \cite{MTsection}. We find that the Jacobian (\ref{Jacobian of specific form in 3.1}) of the specific genus-one fibration we consider here has the term
\begin{equation}
e_4 = f^2,
\end{equation}
which is a perfect square. Owing to this, one may expect that the Jacobian fibration (\ref{Jacobian of specific form in 3.1}) should have two global sections, and thus it has Mordell--Weil rank one (or higher). However, when we study the structure of the global section more carefully, it turns out that this is not the case, and the Jacobian fibration (\ref{Jacobian of specific form in 3.1}) in fact has only one independent global section. 
\par Let us discuss this point. It was described in \cite{MTsection} that when $e_4$ is a perfect square, $e_4=b^2/4$, one can rewrite the equation (\ref{double cover in 3.1}) as
\begin{equation}
\label{double cover when perfect square in 3.1}
(\tau+\frac{1}{2}b)( \tau-\frac{1}{2}b)=\lambda\, (e_0\lambda^3+e_1\lambda^2+e_2\lambda+e_3),
\end{equation}
and $\lambda=\tau\pm \frac{1}{2}b=0$ yields two global sections. This holds for generic coefficients $e_0, e_1, e_2, e_3$, but when $e_3$ vanishes, $e_3=0$, then these no longer yield two global sections. This is because, when $e_3=0$, equation (\ref{double cover when perfect square in 3.1}) becomes 
\begin{equation}
\label{double cover e3 is 0 in 3.1}
(\tau+\frac{1}{2}b)( \tau-\frac{1}{2}b)=\lambda^2\, (e_0\lambda^2+e_1\lambda+e_2),
\end{equation}
and the right-hand side of (\ref{double cover e3 is 0 in 3.1}) has the factor $\lambda^2$, instead of just $\lambda$ as in (\ref{double cover when perfect square in 3.1}). 
\par When $e_3\ne 0$, the ``horizontal'' divisor, namely a global section, increases and the Mordell--Weil rank increases to one \cite{MTsection}, but when $e_3=0$, the ``vertical'' divisor increases instead, and the fibration (\ref{double cover e3 is 0 in 3.1}) acquires a type $I_2$ fiber \footnote{This situation is similar to the limit $b \rightarrow 0$ discussed in \cite{MTsection}. In the limit at which $b\rightarrow 0$, two global sections together form a type $I_2$ fiber (and these no longer yield two sections) \cite{MTsection}. For the limit $b\rightarrow 0$, the left-hand side of the double cover (\ref{double cover when perfect square in 3.1}) becomes a square $\tau^2$, whereas in the case we considered, $e_3=0$, the right-hand side contains a square factor $\lambda^2$.}. 
\par A study of the discriminant of the genus-one fibration also suggests this. Let us provide a demonstration. The discriminant of the double cover of quartic polynomial with $e_3=0$, $e_4=b^2/4$,
\begin{equation}
\tau^2 = e_0\, \lambda^4+e_1\, \lambda^3+e_2\, \lambda^2+b^2/4
\end{equation}
is given as 
\begin{eqnarray}
\label{discriminant when e3 is 0 in 3.1}
\Delta & \sim -\frac{27}{16}\, b^4 e_1^4+9\, e_0 e_2 e_1^2 b^4 -e_2^3 e_1^2 b^2 +4 e_0^3 b^6 - 8\, e_0^2 e_2^2 b^4 +4 e_0 e_2^4 b^2 \\ \nonumber
& = b^2\, \big(-\frac{27}{16}\, b^2 e_1^4+9\, e_0 e_2 e_1^2 b^2 -e_2^3 e_1^2 +4 e_0^3 b^4 - 8\, e_0^2 e_2^2 b^2 +4 e_0 e_2^4 \big).
\end{eqnarray}
The factor $b^2$ in the discriminant (\ref{discriminant when e3 is 0 in 3.1}) represents a type $I_2$ fiber along the divisor $\{b=0\}$ in the base. In our example $e_4=f^2$ and $e_3=0$, and we previously computed that the fibers lying, over the component $\{f=0\}$ have type $I_2$, agreeing with our analysis. 
\par As noted in \cite{MTsection} by a symmetry argument applied to the double cover (\ref{double cover in 3.1}), a bisection splits into a pair of global sections also when $e_0$ is a perfect square (and $e_1\ne 0$). When $e_1=0$ for this situation, similar to what we have just discussed, the double cover instead acquires a type $I_2$ fiber. For our example, $e_0=g^2$, and $e_1=0$. We computed previously that the fibers over the component $\{g=0\}$ have type $I_2$, agreeing with our analysis. 
\par Owing to these arguments, we learn that the Jacobian fibration (\ref{Jacobian of specific form in 3.1}) has only one independent global section and the Mordell--Weil rank is zero \footnote{We would like to remark that, when the specific case where
$$f=t\, \til{f}, \hspace{1cm} g=t\, \til{g},$$
as functions over specific base three-fold $\P^1\times \P^1\times \P^1$ is considered, the complete intersection (\ref{specific complete intersection in 3.1}) yields a genus-one fibered Calabi--Yau four-fold analyzed in \cite{Kdisc}. Similarly, when the specific case $f=g=t$ as functions over $\P^1$ as a base is considered, the complete intersection (\ref{specific complete intersection in 3.1}) yields a genus-one fibered K3 surface analyzed in \cite{K2}. The Jacobian fibrations of these spaces have both Mordell--Weil rank zero \cite{K2, Kdisc}, agreeing with our analysis here.}. If we consider adding variations to the coefficients $a_1, \ldots, a_{10}$, $b_1, \ldots, b_{10}$, of the complete intersection (\ref{specific complete intersection in 3.1}) so the Jacobian has nonzero $e_3$ while keeping the term $e_4=f^2$ in the Jacobian fibration (\ref{Jacobian of specific form in 3.1}) fixed (or the Jacobian has nonzero $e_1$ while keeping the term $e_0=g^2$ fixed) with the bisection geometries condition imposed on the coefficients $a_1, \ldots, a_{10}$, $b_1, \ldots, b_{10}$, we obtain a genus-one fibration with a bisection, the Jacobian fibration of which has two independent sections. Both $U(1)$ and a discrete $\Z_2$ gauge group form in F-theory on this deformed genus-one fibration. The transition from this deformed genus-one fibration to the complete intersection (\ref{specific complete intersection in 3.1}) can be physically seen as the reverse of Higgsing in which an $SU(2)$ gauge group breaks down into $U(1)$ \cite{MTsection}. In this sense, the condition on the parameters yielding the complete intersection (\ref{specific complete intersection in 3.1}) is more restrictive than the deformed genus-one fibration on which both $U(1)$ and a discrete $\Z_2$ gauge group form. 
\par In bisection geometry contained in the four-section geometry, as the locus where a four-section splits into bisections, the complete intersection (\ref{specific complete intersection in 3.1}) and the deformation of this that we have just described have enhanced gauge groups, $SU(2)\times \Z_2$ and $U(1)\times \Z_2$, respectively, unlike a generic member of bisection geometry on which only a discrete gauge group $\Z_2$ forms. If we consider transitions from a generic point in the four-section geometries to these points with enhanced gauge symmetries, the physical viewpoint yields the transitions of gauge groups from $SU(2)\times \Z_2$ or $U(1)\times \Z_2$ to a discrete $\Z_4$ gauge group, which gives a natural Higgsing process. 

\vspace{5mm}

\par If the same argument applies to general situations, not limited to the specific example that we have just discussed, then the physically unnatural phenomenon pointed out in \cite{Kimura1905} can admit a seemingly natural physical interpretation. We find a point in the moduli at which the Mordell--Weil rank of the Jacobian fibration of a genus-one fibration with a bisection increases to one. Namely, we try to find a genus-one fibration with a bisection lacking a global section, the Jacobian fibration of which has two global sections. F-theory compactification on this genus-one fibration has both a discrete $\Z_2$ gauge group and $U(1)$ \footnote{F-theory models on which both a discrete gauge group and $U(1)$ form were considered, e.g., in \cite{KMOPR, GKK}.}. In most situations in 6D F-theory, this $U(1)$ can be un-Higgsed to $SU(2)$ (or higher) without changing the structure of the base space \cite{MTsection}, and the specific example (\ref{specific complete intersection in 3.1}) that we have just discussed precisely corresponds to one of such situations.
\par We choose to interpret that these ``enhanced'' models directly transition to an F-theory model with a discrete $\Z_4$ gauge group in the physical process; then $U(1)\times \Z_2$ (or $SU(2)\times \Z_2$) breaks down into a discrete $\Z_4$ gauge group and no apparent contradiction arises. We expect that, in the four-section geometry when a four-section splits into a pair of bisections, bisection geometries with these enhanced symmetries are likely to be chosen, rather than a generic model with a bisection. 
\par This interpretation seems to resolve the unnatural transition of the gauge groups in the Higgsing process as considered in \cite{Kimura1905}. This interpretation yields a natural transition of the gauge groups in the process. For a four-section geometry (in a specific construction), a study actually showed \cite{Kimura1908} that a discrete $\Z_2$ gauge group is enhanced to $U(1)\times \Z_2$ gauge group along every bisection geometries locus in the moduli of the four-section geometry. 
\par It was shown in \cite{Kimura1908} that the Mordell--Weil rank of the Jacobian fibrations increase to one along every bisection geometries locus in the four-section geometry, where a four-section splits into a pair of bisections, when the four-section geometry is built as the complete intersection of two quadric hypersurfaces in $\P^3$ fibered over a base space. Thus, both $U(1)$ and a discrete $\Z_2$ gauge group arise in F-theory on the bisection geometries loci in the four-section geometry \cite{Kimura1908}. (When double cover ramified along a quartic polynomial is constructed from the two quadric hypersurfaces satisfying (\ref{condition splitting in 2.2}), the coefficient $e_0$ of the term $e_0 \, \lambda^4$ in the quartic polynomial was found to be a perfect square \cite{Kimura1908}. The double cover has the Mordell--Weil rank one \cite{MTsection}, and the double cover yields the Jacobian fibrations of the bisection geometries locus (\ref{condition splitting in 2.2}) \cite{Kimura1908}.) At least for this type of construction of the four-section geometry, an enhancement of the gauge group indeed occurs along any bisection geometries locus. This result in \cite{Kimura1908} supports our solution to the puzzle in \cite{Kimura1905} geometrically. 
\par We consider moving from a generic point in the four-section geometry constructed as the complete intersection of two quadric hypersurfaces in $\P^3$ fibered over a base to any bisection geometries locus, described by the complete intersections of two quadric hypersurfaces with specific coefficients given as (\ref{condition splitting in 2.2}) \footnote{It was demonstrated in \cite{Kimura1908} that every bisection geometries locus in the four-section geometry (\ref{complete intersection general in 2.2}) transforms to (or locus is contained in, after a transformation) the bisection geometries locus of the form (\ref{condition splitting in 2.2}).}. In the string-theoretic language, which reverses the geometric process, the process has the following interpretation: for a generic genus-one fibration with a four-section, one can find some F-theory model with $U(1)\times \Z_2$ gauge group that undergoes transition to F-theory model on that genus-one fibration.   
\par As we remarked previously, in 6D F-theory $U(1)$ arising in F-theory on the genus-one fibration with a bisection can be un-Higgsed to $SU(2)$ in most situations as discussed in \cite{MTsection}.

\subsection{An interpretation of the transition in the trisection model}
\label{sec3.a}
\par We can consider a transition from a generic point in the trisection geometry \cite{KMOPR} to a point (\ref{cubic hypersurface in 2.1}) where a trisection splits into a global section and a bisection. When this is viewed from a physical viewpoint, reversing the geometric order, F-theory model transitions from a model without a gauge group to a model with a discrete $\Z_3$ gauge group via Higgsing, which appears puzzling. 
\par We would like to propose a physically natural interpretation of this puzzle. We can start with a model where a trisection splits into a bisection and a global section given by the equation (\ref{cubic hypersurface in 2.1}), and take a path through the moduli space to a generic trisection model given by the equation (\ref{cubic hypersurface in 2.1}) with added $Y^3$ term with a coefficient. We then consider a deformation of this path, so the bisection in the model that we start with splits into two global sections. As a result, we obtain a path through the moduli wherein a model with a bisection and a global section given by (\ref{cubic hypersurface in 2.1}) transitions to a model with three global sections as an intermediate step, and the model transitions further to a generic trisection model described by (\ref{cubic hypersurface in 2.1}) with added $Y^3$ term with a coefficient. This yields a transition from an F-theory model with two $U(1)$ factors \footnote{An elliptic fibration with three independent global sections has Mordell--Weil rank two, therefore, an F-theory model on that elliptic fibration has two $U(1)$ factors.} breaking down into another model with a discrete $\Z_3$ gauge group via Higgsing, which is natural.  
\par In section \ref{sec3.2}, we generalize this interpretation to a general $n$-section model. 
\par Now we would like to discuss two loci in the moduli (\ref{cubic hypersurface in 2.1}) where $SU(2)$ gauge group forms instead of the two $U(1)$ factors that we discussed previously. Transition from an F-theory model belonging to these loci to a generic trisection model yields another natural physical interpretation.

\par First, we consider a special locus: $c_8=0$ in the geometry of cubic hypersurfaces (\ref{cubic hypersurface in 2.1}) that we considered in section \ref{sec2.1}. The Weierstrass coefficients of the Jacobian becomes as follows along the locus $c_8=0$:
\begin{eqnarray}
\label{cubic hypersurface c8 in 3.a}
f= & 3\cdot \Big[ -\big(c_9^2-4(c_3c_5+c_2c_7)\big)^2 \\ \nonumber
& +24[-c_9 (c_4c_2c_3+c_3c_6c_7) +2(c_3^2c_6c_4+c_2c_3c_5c_7+c_1c_3c_7^2)] \Big]. \\ \nonumber
g= & 2\cdot \Big[ \big(c_9^2-4(c_3c_5+c_2c_7)\big)^3 \\ \nonumber
& -36\big(c_9^2-4(c_3c_5+c_2c_7)\big)\Big(-c_9(c_2c_3c_4+c_3c_6c_7) +2(c_3^2c_6c_4+c_2c_3c_5c_7+c_1c_3c_7^2) \Big) \\ \nonumber
& +216\Big( (c_2c_3c_4+c_3c_6c_7)^2 +4\big( -c_1c_3^3c_4^2+c_3^2c_4(-c_2c_6+c_1c_9)c_7-c_1c_3^2c_5c_7^2 \big) \Big) \Big].
\end{eqnarray}
Computation shows that the discriminant of the Jacobian along the locus $c_8=0$, whose Weierstrass coefficients are given as in (\ref{cubic hypersurface c8 in 3.a}), has factor $c_3^2$:
\begin{equation}
\Delta(c_8=0) \sim c_3^2\cdot \til{\Delta},
\end{equation}
where $\til{\Delta}$ does not have factor $c_3$. From the discriminant (\ref{discriminant c2568 in 2.1}) that we computed in section \ref{sec2.1}, we deduce that $\til{\Delta}$ is either irreducible, or the product of a linear factor (which reduces to $c_1$ when $c_2=c_5=c_6=0$) and an irreducible factor. Therefore, we learn that $SU(2)$ gauge group forms in 6D F-theory on the Jacobian along the locus $c_8=0$. 
\par Cubic hypersurface (\ref{cubic hypersurface in 2.1}) has a constant section as we mentioned in section \ref{sec2.1}. Along the locus $c_8=0$, the cubic hypersurface (\ref{cubic hypersurface in 2.1}) becomes as follows when we set $X=0$:
\begin{equation}
\label{enhancement equation in 3.a}
Z^2 (c_4\, Z+ c_7\, Y) =0.
\end{equation}
From the equation (\ref{enhancement equation in 3.a}), at first sight, it appears that the hypersurface admits a free section, $[X:Y:Z]=[0:c_4:-c_7]$, along the locus $c_8=0$; however, this is not the case, and the hypersurface does not admit a free section along the locus $c_8=0$. This is because the equation (\ref{enhancement equation in 3.a}) has the square factor $Z^2$. If one adds some deformation to perturb the term $Z^2$ to a non-square degree-two term (while we keep the coefficient of $Y^3$ in the defining equation of the cubic hypersurface zero), the hypersurface admits a free section and the bisection splits into two global sections. Because the cubic hypersurface (\ref{cubic hypersurface in 2.1}) has a constant section $[X:Y:Z]=[0:1:0]$, when the bisection splits into two global sections, the hypersurface acquires three global sections, and the Mordell--Weil rank is enhanced to two; $U(1)^2$ forms in F-theory on this deformed hypersurface away from the locus $c_8=0$. An $SU(2)$ gauge group, instead of $U(1)^2$, forms along the locus $c_8=0$ owing to a mechanism similar to that discussed in \cite{MTsection}. 
\par The Jacobian (\ref{cubic hypersurface c8 in 3.a}) of cubic hypersurface along the locus $c_8=0$ corresponds to the bisection model \cite{BM, MTsection} constructed as double cover of the quartic polynomial $\tau^2=e_0 \lambda^4+ e_1 \lambda^3+ e_2 \lambda^2+ e_3 \lambda + e_4$, when the parameters $e_0, e_1, e_2, e_3$, and $e_4$ assume the following values:
\begin{eqnarray}
\label{coeffs c8 in 3.a}
e_0 = & -432\, [(c_2c_4+c_6c_7)^2+4\big(-c_1c_3c_4^2+c_4(-c_2c_6+c_1c_9)c_7-c_1c_5c_7^2\big)] \\ \nonumber
e_1 = & 72 \, [-c_9(c_2c_4+c_6c_7)+2(c_3c_4c_6+c_2c_5c_7+c_1c_7^2)] \\ \nonumber
e_2 = & -3\big(c_9^2-4(c_3c_5+c_2c_7)\big) \\ \nonumber
e_3= & c_3 \\ \nonumber
e_4 = & 0.
\end{eqnarray}
We note that the coefficient $e_4$ vanishes in (\ref{coeffs c8 in 3.a}). The models along the locus $c_8=0$ correspond to a limit at which the coefficient $e_4$ goes to zero.
\par The cubic hypersurface (\ref{cubic hypersurface in 2.1}) is symmetric under the exchange of $X$ and $Z$, when one considers the following exchange of the coefficients:
\begin{eqnarray}
c_1 \longleftrightarrow & c_4 \\ \nonumber
c_2 \longleftrightarrow & c_7 \\ \nonumber
c_3 \longleftrightarrow & c_8 \\ \nonumber
c_5 \longleftrightarrow & c_6.
\end{eqnarray}
Physics of F-theory on the cubic hypersurface (\ref{cubic hypersurface in 2.1}) is invariant under the exchange. Therefore, F-theory models along the locus $c_3=0$ are identical to those along $c_8=0$. The discriminant of the Jacobian along $c_3=0$ contains square factor $c_8^2$, and an $SU(2)$ gauge group forms in F-theory along $c_3=0$.

\par A trisection splits into a global section and a bisection for cubic hypersurface (\ref{cubic hypersurface in 2.1}) where the coefficient of $Y^3$ vanishes, and the bisection further splits into two global sections in a certain limit \cite{MTsection}. This yields a point in the moduli where a trisection splits into three global sections, and $U(1)^2$ forms in F-theory in this limit. F-theory models along the two loci, $c_3=0$ and $c_8=0$, have an $SU(2)$ gauge group, instead of $U(1)^2$. One can consider the transition from an F-theory model along one of the two loci, $c_3=0$ and $c_8=0$ in the cubic hypersurfaces (\ref{cubic hypersurface in 2.1}) to the generic trisection model. $SU(2)$ breaks down into a discrete $\Z_3$ gauge group via Higgsing in the transition, yielding another physically natural process.

\vspace{5mm}

\par Based on the arguments we have made in sections \ref{sec3.1} and \ref{sec3.a}, in section \ref{sec3.2} we discuss a possible physical interpretation of the phenomenon we discussed in section \ref{sec2.1}.

\subsection{An interpretation of the transition from a model without $U(1)$ or discrete gauge group to a model with a discrete $\Z_n$ gauge group}
\label{sec3.2}
\par By using a physical interpretation of the unnatural phenomenon pointed out in \cite{Kimura1905} that we proposed in section \ref{sec3.1} and a physical interpretation of the unnatural phenomenon observed in the trisection geometry that we proposed in section \ref{sec3.a}, we discuss a possible interpretation of the puzzle we considered in section \ref{sec2.1}. As observed in section \ref{sec2.1}, in the multisection geometry, an F-theory model without a gauge group transitions to another model with a discrete $\Z_n$ gauge group via Higgsing \footnote{As we stated in the introduction and section \ref{sec2.1}, the puzzling physical phenomenon that we discussed in section \ref{sec2.1} occurs when $n\ge 3$. We only consider the situations $n\ge 3$ here.}. What appears puzzling here is that, although there is supposed to be some gauge group that is breaking down into the discrete $\Z_n$ gauge group via Higgsing, the discrete $\Z_n$ gauge group rather appears to arise ``from nothing.'' If there appears to be no gauge group that breaks down into the discrete $\Z_n$ gauge group, how should we interpret this process?
\par A natural physical interpretation might be that, similar to what we proposed in section \ref{sec3.1}, the model corresponding to the point in the moduli of $n$-section geometry at which an $n$-section splits into a global section and an $(n-1)$-section deforms to another model with an enhanced gauge symmetry, such as a non-Abelian gauge group or $U(1)$. If so, this allows us to view the process as a non-Abelian gauge group or $U(1)$ breaking down into a discrete $\Z_n$ gauge group, which is natural. 
\par Given a general multisection (of degree greater than two), the splitting processes of the multisection into a global section and a multisection as an intermediate step, before it finally splits into multiple sheets of global sections, are prevalent in the multisection geometry. Therefore, the puzzling process of an F-theory model without having a gauge group undergoing transition to another model with a discrete $\Z_n$ gauge group via a Higgsing process inevitably appears in the moduli. Some physical interpretation, such as the one we proposed, seems to be required as a ``way out'' to avoid this puzzling transition from no gauge symmetry to a discrete $\Z_n$ gauge symmetry via Higgsing.
\par The physical interpretation that we proposed is possible when a model with a single global section (with an $(n-1)$-section) in the moduli of an $n$-section admits a deformation to another model with two global sections, or a model with a non-Abelian gauge group such as $SU(2)$. This deformation is, in fact, possible for generic 6D F-theory models. Let us provide a sketch of a proof of this. 
\par Similar to the process in which an $n$-section splits into a global section and an $(n-1)$-section, one can consider the iteration where the resulting $(n-1)$-section splits into a global section and an $(n-2)$-section. Putting the two steps together, one can view it as an $n$-section splitting into two global sections and an $(n-2)$-section. The $U(1)$ gauge group arises in F-theory on the resulting elliptic fibration with two global sections. As discussed in \cite{MTsection}, in 6D F-theory this $U(1)$ can be further un-Higgsed to $SU(2)$ in most situations.
\par The two global sections obtained here are generically independent, and thus they generate a rank-one Mordell--Weil group. We would like to demonstrate this point here: We can also consider a process wherein an $n$-section splits into a bisection and an $(n-2)$-section, and the bisection further splits into two global sections, instead of an $n$-section splitting into a global section and an $(n-1)$-section, and the $(n-1)$-section further splitting into an $(n-2)$-section and another global section. The process of a bisection splitting into two global sections was analyzed in \cite{MTsection}. The two global sections into which a bisection is split are independent and they generate Mordell--Weil group of rank one \cite{MTsection}; therefore, the elliptic fibrations corresponding to the generic points in the moduli of the $n$-section geometry at which an $n$-section splits into two global sections and an $(n-2)$-section have the Mordell--Weil rank one.
\par The case $n=3$ precisely corresponds to an interpretation that we proposed in section \ref{sec3.a}, wherein a trisection splits into a global section and a bisection, then the bisection further splits into two global sections. For this special case, the resulting elliptic fibration has Mordell--Weil rank two as we mentioned previously, while for larger $n$, $n>3$, the resulting elliptic fibration has only two independent global sections (and an $(n-2)$-section), and the Mordell--Weil rank is one.
\par Briefly, given a locus in the $n$-section moduli at which an $n$-section splits into a global section and an $(n-1)$-section, a point with two global sections or a point where $U(1)$ is further enhanced to $SU(2)$ can be found. We consider a transition from a generic point in the moduli with an $n$-section to such points with enhanced gauge symmetries. The physical viewpoint of this reverses the order, and $U(1)$ or $SU(2)$ breaks down into a discrete $\Z_n$, yielding a natural Higgsing process of gauge groups. We take this as a physical interpretation of the puzzling phenomenon we observed in section \ref{sec2.1}. This at least provides one natural interpretation of the phenomenon.
\par The interpretations of the puzzle that we proposed in this study is one possibility that can resolve the puzzle that we defined in section \ref{sec2.1}. Determining whether our interpretations yield a solution to the puzzle is left for future studies. There may be other possible solution(s) to the puzzle. 

\section{Open problems and some remarks}
\label{sec4}
\par In this note, we have pointed out a new puzzle, which is different from the unnatural phenomenon observed in \cite{Kimura1905}, by analyzing the splitting process of a multisection into a global section and a multisection of smaller degree; in other words, we observed a phenomenon in which an F-theory model with no $U(1)$ and no discrete gauge group (and no non-Abelian gauge group) transitions to another F-theory model with a discrete $\Z_n$ gauge group via Higgsing. Because the original model in this transition does not have a gauge symmetry, a discrete gauge group appears to arise ``from nothing'' in the Higgsing process, and this appears physically puzzling. 
\par We also proposed possible interpretations that can resolve the new puzzle. There are various ways in which an $n$-section splits into multisections of smaller degrees. Additionally, we proposed an interpretation that can resolve the puzzle observed in \cite{Kimura1905}. Does our interpretation suggest that when multisection splits, a specific way of splitting of an $n$-section in the $n$-section geometry is chosen so the corresponding F-theory model has an enhanced gauge symmetry? It might be interesting to study whether there is any reason that other ways of splitting of multisections are ruled out owing to some physical mechanism, or whether there is a reason that the ways of splitting in which the corresponding model has an enhanced gauge symmetry are favored over other possibilities. This will be a likely direction of future study.  
\par However, one of the central motivations of this note was to point out a new puzzle as we defined in section \ref{sec2.1}, and whether the interpretations that we proposed in this work actually yield a solution to the puzzle is left undetermined.
\par The puzzling physical phenomena we observed in this study and observed in \cite{Kimura1905}, at the level of geometry, do not depend on the dimensions. However, we only considered 6D F-theory when we proposed physical interpretations of these phenomena in this note. This is owing to the issue of flux for 4D F-theory: the superpotential generated by flux may alter the arguments that worked without the insertion of a flux, as noted in \cite{MTsection}. Meanwhile, the effect of inserting a flux may explain our proposal. It might be interesting to consider this possibility, and this is also a likely direction of future study. 
\par The situation in which a four-section splits into a pair of bisections generalizes to those of multisections of higher degrees. For example, when a multisection has degree a multiple of two, say $2n$, we expect that there is a situation in which the $2n$-section splits into a pair of $n$-sections. A multisection of degree $3n$ splitting into a triplet of $n$-sections yields another example. From physical viewpoint, an F-theory model with a discrete $\Z_n$ gauge group transitions to another model with a discrete $\Z_{2n}$ gauge group, and an F-theory model with a discrete $\Z_n$ gauge group undergoes transition to another model with a discrete $\Z_{3n}$ gauge group, via Higgsing, respectively. These processes also appear puzzling. Studying these is also a likely target of future study.

\section*{Acknowledgments}

We would like to thank Shun'ya Mizoguchi and Shigeru Mukai for discussions.


\begin{thebibliography}{99}

\bibitem{Vaf}C.~Vafa, ``Evidence for F-theory'', {\it Nucl. Phys.} {\bf B 469} (1996) 403 [arXiv:hep-th/9602022].
\bibitem{MV1}D.~R.~Morrison and C.~Vafa, ``Compactifications of F-theory on Calabi-Yau threefolds. 1'', {\it Nucl. Phys.} {\bf B 473} (1996) 74 [arXiv:hep-th/9602114].
\bibitem{MV2}D.~R.~Morrison and C.~Vafa, ``Compactifications of F-theory on Calabi-Yau threefolds. 2'', {\it Nucl. Phys.} {\bf B 476} (1996) 437 [arXiv:hep-th/9603161].

\bibitem{MorrisonPark}D.~R.~Morrison and D.~S.~Park, ``F-Theory and the Mordell-Weil Group of Elliptically-Fibered Calabi-Yau Threefolds'', {\it JHEP} {\bf 10} (2012) 128 [arXiv:1208.2695 [hep-th]].
\bibitem{MPW}C.~Mayrhofer, E.~Palti and T.~Weigand, ``U(1) symmetries in F-theory GUTs with multiple sections'', {\it JHEP} {\bf 03} (2013) 098 [arXiv:1211.6742 [hep-th]].
\bibitem{BGK}V.~Braun, T.~W.~Grimm and J.~Keitel, ``New Global F-theory GUTs with U(1) symmetries'', {\it JHEP} {\bf 09} (2013) 154 [arXiv:1302.1854 [hep-th]].
\bibitem{BMPWsection}J.~Borchmann, C.~Mayrhofer, E.~Palti and T.~Weigand, ``Elliptic fibrations for $SU(5)\times U(1)\times U(1)$ F-theory vacua'', {\it Phys. Rev.} {\bf D88} (2013) no.4 046005 [arXiv:1303.5054 [hep-th]].
\bibitem{CKP}M.~Cveti\v c, D.~Klevers and H.~Piragua, ``F-Theory Compactifications with Multiple U(1)-Factors: Constructing Elliptic Fibrations with Rational Sections'', {\it JHEP} {\bf 06} (2013) 067 [arXiv:1303.6970 [hep-th]].
\bibitem{BGK1306}V.~Braun, T.~W.~Grimm and J.~Keitel, ``Geometric Engineering in Toric F-Theory and GUTs with U(1) Gauge Factors,'' {\it JHEP} {\bf 12} (2013) 069 [arXiv:1306.0577 [hep-th]].
\bibitem{CGKP}M.~Cveti\v c, A.~Grassi, D.~Klevers and H.~Piragua, ``Chiral Four-Dimensional F-Theory Compactifications With SU(5) and Multiple U(1)-Factors'', {\it JHEP} {\bf 04} (2014) 010 [arXiv:1306.3987 [hep-th]].
\bibitem{CKP1307}M.~Cveti\v c, D.~Klevers and H.~Piragua, ``F-Theory Compactifications with Multiple U(1)-Factors: Addendum'', {\it JHEP} {\bf 12} (2013) 056 [arXiv:1307.6425 [hep-th]].
\bibitem{CKPS}M.~Cveti\v c, D.~Klevers, H.~Piragua and P.~Song, ``Elliptic fibrations with rank three Mordell-Weil group: F-theory with U(1) x U(1) x U(1) gauge symmetry,'' {\it JHEP} {\bf 1403} (2014) 021 [arXiv:1310.0463 [hep-th]].
\bibitem{AL}I.~Antoniadis and G.~K.~Leontaris, ``F-GUTs with Mordell-Weil U(1)'s,'' {\it Phys. Lett.} {\bf B735} (2014) 226--230 [arXiv:1404.6720 [hep-th]].
\bibitem{EKY1410}M.~Esole, M.~J.~Kang and S.-T.~Yau, ``A New Model for Elliptic Fibrations with a Rank One Mordell-Weil Group: I. Singular Fibers and Semi-Stable Degenerations'', [arXiv:1410.0003 [hep-th]].
\bibitem{LSW}C.~Lawrie, S.~Sch\"afer-Nameki and J.-M.~Wong, ``F-theory and All Things Rational: Surveying U(1) Symmetries with Rational Sections'', {\it JHEP} {\bf 09} (2015) 144 [arXiv:1504.05593 [hep-th]]. 
\bibitem{CKPT}M.~Cveti\v c, D.~Klevers, H.~Piragua and W.~Taylor, ``General U(1)$\times$U(1) F-theory compactifications and beyond: geometry of unHiggsings and novel matter structure,'' {\it JHEP} {\bf 1511} (2015) 204 [arXiv:1507.05954 [hep-th]].
\bibitem{CGKPS}M.~Cveti\v c, A.~Grassi, D.~Klevers, M.~Poretschkin and P.~Song, ``Origin of Abelian Gauge Symmetries in Heterotic/F-theory Duality,'' {\it JHEP} {\bf 1604} (2016) 041 [arXiv:1511.08208 [hep-th]].
\bibitem{MP2}D.~R.~Morrison and D.~S.~Park, ``Tall sections from non-minimal transformations'', {\it JHEP} {\bf 10} (2016) 033 [arXiv:1606.07444 [hep-th]].
\bibitem{MPT1610}D.~R.~Morrison, D.~S.~Park and W.~Taylor, ``Non-Higgsable abelian gauge symmetry and $\mathrm{F}$-theory on fiber products of rational elliptic surfaces'', {\it Adv. Theor. Math. Phys.} {\bf 22} (2018) 177--245 [arXiv:1610.06929 [hep-th]].
\bibitem{BMW2017}M.~Bies, C.~Mayrhofer and T.~Weigand, ``Gauge Backgrounds and Zero-Mode Counting in F-Theory'', {\it JHEP} {\bf 11} (2017) 081 [arXiv:1706.04616 [hep-th]].
\bibitem{CL2017}M.~Cveti\v c and L.~Lin, ``The Global Gauge Group Structure of F-theory Compactification with U(1)s'', {\it JHEP} {\bf 01} (2018) 157 [arXiv:1706.08521 [hep-th]].
\bibitem{BMW1706}M.~Bies, C.~Mayrhofer and T.~Weigand, ``Algebraic Cycles and Local Anomalies in F-Theory'', {\it JHEP} {\bf 11} (2017) 100 [arXiv:1706.08528 [hep-th]].
\bibitem{KimuraMizoguchi}Y.~Kimura and S.~Mizoguchi, ``Enhancements in F-theory models on moduli spaces of K3 surfaces with $ADE$ rank 17'', {\it PTEP} {\bf 2018} no. 4 (2018) 043B05 [arXiv:1712.08539 [hep-th]].
\bibitem{Kimura1802}Y.~Kimura, ``F-theory models on K3 surfaces with various Mordell-Weil ranks -constructions that use quadratic base change of rational elliptic surfaces'', {\it JHEP} {\bf 05} (2018) 048 [arXiv:1802.05195 [hep-th]].
\bibitem{LRW2018}S.-J.~Lee, D.~Regalado and T.~Weigand, ``6d SCFTs and U(1) Flavour Symmetries'', {\it JHEP} {\bf 11} (2018) 147 [arXiv:1803.07998 [hep-th]].
\bibitem{MizTani2018}S.~Mizoguchi and T.~Tani, ``Non-Cartan Mordell-Weil lattices of rational elliptic surfaces and heterotic/F-theory compactifications'', {\it JHEP} {\bf 03} (2019) 121 [arXiv:1808.08001 [hep-th]].
\bibitem{CMPV1811}F.~M.~Cianci, D.~K.~Mayorga Pena and R.~Valandro, ``High U(1) charges in type IIB models and their F-theory lift'', {\it JHEP} {\bf 04} (2019) 012 [arXiv:1811.11777 [hep-th]].
\bibitem{TT2019}W.~Taylor and A.~P.~Turner, ``Generic matter representations in 6D supergravity theories'', {\it JHEP} {\bf 05} (2019) 081 [arXiv:1901.02012 [hep-th]].
\bibitem{Kimura1903}Y.~Kimura, ``F-theory models with 3 to 8 U(1) factors on K3 surfaces'' [arXiv:1903.03608 [hep-th]].
\bibitem{EJ1905}M.~Esole and P.~Jefferson, ``The Geometry of SO(3), SO(5), and SO(6) models'' [arXiv:1905.12620 [hep-th]].
\bibitem{LW1905}S.-J.~Lee and T.~Weigand, ``Swampland Bounds on the Abelian Gauge Sector'', {\it Phys. Rev.} {\bf D100} (2019) no.2 026015 [arXiv:1905.13213 [hep-th]].

\bibitem{KNPRR}T.~Kobayashi, H.~P.~Nilles, F.~Ploger, S.~Raby and M.~Ratz, ``Stringy origin of non-Abelian discrete flavor symmetries,'' {\it Nucl.Phys.} {\bf B768} (2007) 135 [arXiv: hep-ph/0611020].
\bibitem{ACKO} H.~Abe, K.-S.~Choi, T.~Kobayashi and H.~Ohki, ``Non-Abelian Discrete Flavor Symmetries from Magnetized/Intersecting Brane Models,'' {\it Nucl.Phys.} {\bf B820} (2009) 317 [arXiv:0904.2631 [hep-ph]].
\bibitem{BS}T.~Banks and N.~Seiberg, ``Symmetries and Strings in Field Theory and Gravity'', {\it Phys. Rev.} {\bf D83} (2011) 084019 [arXiv:1011.5120 [hep-th]].
\bibitem{HSsums}S.~Hellerman and E.~Sharpe, ``Sums over topological sectors and quantization of Fayet-Iliopoulos parameters'', {\it Adv.Theor.Math.Phys.} {\bf 15} (2011) 1141--1199 [arXiv:1012.5999 [hep-th]].
\bibitem{CIM}P.~G.~Camara, L.~E.~Ibanez and F.~Marchesano, ``RR photons'', {\it JHEP} {\bf 09} (2011) 110 [arXiv:1106.0060 [hep-th]].
\bibitem{BISU}M.~Berasaluce-Gonzalez, L.~E.~Ibanez, P.~Soler and A.~M.~Uranga, ``Discrete gauge symmetries in D-brane models'', {\it JHEP} {\bf 12} (2011) 113 [arXiv:1106.4169 [hep-th]].
\bibitem{ISU}L.~E.~Ibanez, A.~N.~Schellekens and A.~M.~Uranga, ``Discrete Gauge Symmetries in Discrete MSSM-like Orientifolds'', {\it Nucl.Phys.} {\bf B865} (2012) 509--540 [arXiv:1205.5364 [hep-th]].
\bibitem{BCMRU}M.~Berasaluce-Gonzalez, P.~G.~Camara, F.~Marchesano, D.~Regalado and A.~M.~Uranga, ``Non-Abelian discrete gauge symmetries in 4d string models'', {\it JHEP} {\bf 09} (2012) 059 [arXiv:1206.2383 [hep-th]].
\bibitem{BCMU}M.~Berasaluce-Gonzalez, P.~G.~Camara, F.~Marchesano and A.~M.~Uranga, ``Zp charged branes in flux compactifications'', {\it JHEP} {\bf 04} (2013) 138 [arXiv:1211.5317 [hep-th]]. 
\bibitem{MRV}F.~Marchesano, D.~Regalado and L.~Vazquez-Mercado, ``Discrete flavor symmetries in D-brane models'', {\it JHEP} {\bf 09} (2013) 028 [arXiv:1306.1284 [hep-th]].
\bibitem{HS}G.~Honecker and W.~Staessens, ``To Tilt or Not To Tilt: Discrete Gauge Symmetries in Global Intersecting D-Brane Models'', {\it JHEP} {\bf 10} (2013) 146 [arXiv:1303.4415 [hep-th]].
\bibitem{BRU}M.~Berasaluce-Gonzalez, G.~Ramirez and A.~M.~Uranga, ``Antisymmetric tensor $Z_p$ gauge symmetries in field theory and string theory'', {\it JHEP} {\bf 01} (2014) 059 [arXiv:1310.5582 [hep-th]].
\bibitem{KKLM}A.~Karozas, S.~F.~King, G.~K.~Leontaris and A.~Meadowcroft, ``Discrete Family Symmetry from F-Theory GUTs'', {\it JHEP} {\bf 09} (2014) 107 [arXiv:1406.6290 [hep-ph]]. 
\bibitem{HS2}G.~Honecker and W.~Staessens, ``Discrete Abelian gauge symmetries and axions'', {\it J.Phys.Conf.Ser.} {\bf 631} (2015) no.1 [arXiv:1502.00985 [hep-th]].
\bibitem{GPR}T.~W.~Grimm, T.~G.~Pugh and D.~Regalado, ``Non-Abelian discrete gauge symmetries in F-theory'', {\it JHEP} {\bf 02} (2016) 066 [arXiv:1504.06272 [hep-th]]. 

\bibitem{MTsection}D.~R.~Morrison and W.~Taylor, ``Sections, multisections, and $U(1)$ fields in F-theory'', {\it J. Singularities} {\bf 15} (2016) 126--149 [arXiv:1404.1527 [hep-th]].

\bibitem{BDHKMMS}J.~de~Boer, R.~Dijkgraaf, K.~Hori, A.~Keurentjes, J.~Morgan, D.~R.~Morrison and S.~Sethi, ``Triples, fluxes, and strings'', {\it Adv. Theor. Math. Phys.} {\bf 4} (2002) 995--1186 [arXiv: hep-th/0103170].

\bibitem{BM}V.~Braun and D.~R.~Morrison, ``F-theory on Genus-One Fibrations'', {\it JHEP} {\bf 08} (2014) 132 [arXiv:1401.7844 [hep-th]].
\bibitem{AGGK}L.~B.~Anderson, I.~Garcia-Etxebarria, T.~W.~Grimm and J.~Keitel, ``Physics of F-theory compactifications without section'', {\it JHEP} {\bf 12} (2014) 156 [arXiv:1406.5180 [hep-th]].
\bibitem{KMOPR}D.~Klevers, D.~K.~Mayorga Pena, P.~K.~Oehlmann, H.~Piragua and J.~Reuter, ``F-Theory on all Toric Hypersurface Fibrations and its Higgs Branches'', {\it JHEP} {\bf 01} (2015) 142 [arXiv:1408.4808 [hep-th]].
\bibitem{GGK}I.~Garcia-Etxebarria, T.~W.~Grimm and J.~Keitel, ``Yukawas and discrete symmetries in F-theory compactifications without section'', {\it JHEP} {\bf 11} (2014) 125 [arXiv:1408.6448 [hep-th]].
\bibitem{MPTW}C.~Mayrhofer, E.~Palti, O.~Till and T.~Weigand, ``Discrete Gauge Symmetries by Higgsing in four-dimensional F-Theory Compactifications'', {\it JHEP} {\bf 12} (2014) 068 [arXiv:1408.6831 [hep-th]].
\bibitem{MPTW2}C.~Mayrhofer, E.~Palti, O.~Till and T.~Weigand, ``On Discrete Symmetries and Torsion Homology in F-Theory'', {\it JHEP} {\bf 06} (2015) 029 [arXiv:1410.7814 [hep-th]].
\bibitem{BGKintfiber}V.~Braun, T.~W.~Grimm and J.~Keitel, ``Complete Intersection Fibers in F-Theory'', {\it JHEP} {\bf 03} (2015) 125 [arXiv:1411.2615 [hep-th]].
\bibitem{CDKPP}M.~Cveti\v c, R.~Donagi, D.~Klevers, H.~Piragua and M.~Poretschkin, ``F-theory vacua with $\mathbb Z_3$ gauge symmetry'', {\it Nucl. Phys.} {\bf B898} (2015) 736--750 [arXiv:1502.06953 [hep-th]].
\bibitem{LMTW}L.~Lin, C.~Mayrhofer, O.~Till and T.~Weigand, ``Fluxes in F-theory Compactifications on Genus-One Fibrations'', {\it JHEP} {\bf 01} (2016) 098 [arXiv:1508.00162 [hep-th]].
\bibitem{K}Y.~Kimura, ``Gauge Groups and Matter Fields on Some Models of F-theory without Section'', {\it JHEP} {\bf 03} (2016) 042 [arXiv:1511.06912 [hep-th]].
\bibitem{K2}Y.~Kimura, ``Gauge symmetries and matter fields in F-theory models without section- compactifications on double cover and Fermat quartic K3 constructions times K3'', {\it Adv. Theor. Math. Phys.} {\bf 21} (2017) no.8, 2087--2114 [arXiv:1603.03212 [hep-th]].
\bibitem{ORS1604}P.-K.~Oehlmann, J.~Reuter and T.~Schimannek, ``Mordell-Weil Torsion in the Mirror of Multi-Sections'', {\it JHEP} {\bf 12} (2016) 031 [arXiv:1604.00011 [hep-th]].
\bibitem{KCY4}Y.~Kimura, ``Gauge groups and matter spectra in F-theory compactifications on genus-one fibered Calabi-Yau 4-folds without section - Hypersurface and double cover constructions'', {\it Adv. Theor. Math. Phys.} {\bf 22} (2018) no.6, 1489--1533 [arXiv:1607.02978 [hep-th]]. 
\bibitem{CGP}M.~Cveti\v c, A.~Grassi and M.~Poretschkin, ``Discrete Symmetries in Heterotic/F-theory Duality and Mirror Symmetry,'', {\it JHEP} {\bf 06} (2017) 156 [arXiv:1607.03176 [hep-th]]. 
\bibitem{Kdisc}Y.~Kimura, ``Discrete Gauge Groups in F-theory Models on Genus-One Fibered Calabi-Yau 4-folds without Section'', {\it JHEP} {\bf 04} (2017) 168 [arXiv:1608.07219 [hep-th]].
\bibitem{Kimura1801}Y.~Kimura, ``K3 surfaces without section as double covers of Halphen surfaces, and F-theory compactifications'', {\it PTEP} {\bf 2018} (2018) 043B06 [arXiv:1801.06525 [hep-th]].
\bibitem{AGGO1801}L.~B.~Anderson, A.~Grassi, J.~Gray and P.-K.~Oehlmann, ``F-theory on Quotient Threefolds with (2,0) Discrete Superconformal Matter'', {\it JHEP} {\bf 06} (2018) 098 [arXiv:1801.08658 [hep-th]].
\bibitem{Kimura1806}Y.~Kimura, ``$SU(n) \times \mathbb{Z}_2$ in F-theory on K3 surfaces without section as double covers of Halphen surfaces'' [arXiv:1806.01727 [hep-th]].
\bibitem{TasilectWeigand}T.~Weigand, ``F-theory'', {\it PoS} {\bf TASI2017} (2018) 016 [arXiv:1806.01854 [hep-th]].
\bibitem{CLLO}M.~Cveti\v c, L.~Lin, M.~Liu and P.-K.~Oehlmann, ``An F-theory Realization of the Chiral MSSM with $\mathbb{Z}_2$-Parity'', {\it JHEP} {\bf 09} (2018) 089 [arXiv:1807.01320 [hep-th]]
\bibitem{TasilectCL}M.~Cveti\v c and L.~Lin, ``TASI Lectures on Abelian and Discrete Symmetries in F-theory'', {\it PoS} {\bf TASI2017} (2018) 020 [arXiv:1809.00012 [hep-th]].
\bibitem{HT}Y.-C.~Huang and W.~Taylor, ``On the prevalence of elliptic and genus one fibrations among toric hypersurface Calabi-Yau threefolds'', {\it JHEP} {\bf 03} (2019) 014 [arXiv:1809.05160 [hep-th]].
\bibitem{Kimura1810}Y.~Kimura, ``Nongeometric heterotic strings and dual F-theory with enhanced gauge groups'', {\it JHEP} {\bf 02} (2019) 036 [arXiv:1810.07657 [hep-th]].
\bibitem{Kimura1902}Y.~Kimura, ``Unbroken $E_7\times E_7$ nongeometric heterotic strings, stable degenerations and enhanced gauge groups in F-theory duals'' [arXiv:1902.00944 [hep-th]].
\bibitem{Kimura1905}Y.~Kimura, ``Discrete gauge groups in certain F-theory models in six dimensions'', {\it JHEP} {\bf 07} (2019) 027 [arXiv:1905.03775 [hep-th]].

\bibitem{BEFNQ}P.~Berglund, J.~Ellis, A.~E.~Faraggi, D.~V.~Nanopoulos and Z.~Qiu, ``Elevating the free fermion $Z_2\times Z_2$ orbifold model to a compactification of F-theory'', {\it Int. Jour. of Mod. Phys.} {\bf A 15} (2000) 1345--1362 [arXiv:hep-th/9812141].

\bibitem{Kimura1908}Y.~Kimura, ``F-theory models with $U(1)\times \mathbb{Z}_2,\,  \mathbb{Z}_4$ and transitions in discrete gauge groups'', {\it JHEP} {\bf 03} (2020) 153 [arXiv:1908.06621 [hep-th]].

\bibitem{BB}K.~Becker and M.~Becker, ``M theory on eight manifolds'', {\it Nucl. Phys.} {\bf B477} (1996) 155--167 [arXiv: hep-th/9605053].
\bibitem{SVW}S.~Sethi, C.~Vafa and E.~Witten, ``Constraints on low dimensional string compactifications'', {\it Nucl. Phys.} {\bf B 480} (1996) 213--224, [arXiv: hep-th/9606122].
\bibitem{W96}E.~Witten, ``On flux quantization in M theory and the effective action'', {\it J. Geom. Phys.} {\bf 22} (1997) 1--13 [arXiv: hep-th/9609122].
\bibitem{GVW}S.~Gukov, C.~Vafa and E.~Witten, ``CFT's from Calabi-Yau four folds'', {\it Nucl. Phys.} {\bf B584} (2000) 69--108 [arXiv: hep-th/9906070].
\bibitem{DRS}K.~Dasgupta, G.~Rajesh and S.~Sethi, ``M theory, orientifolds and G-flux'', {\it JHEP} {\bf 08} (1999) 023 [arXiv: hep-th/9908088].

\bibitem{MSSN}J.~Marsano, N.~Saulina and S.~Sch\"afer-Nameki, ``	
A Note on G-Fluxes for F-theory Model Building'', {\it JHEP} {\bf 11} (2010) 088 [arXiv:1006.0483 [hep-th]].
\bibitem{CS}A.~Collinucci and R.~Savelli, ``On Flux Quantization in F-Theory'', {\it JHEP} {\bf 02} (2012) 015 [arXiv:1011.6388 [hep-th]].
\bibitem{MSSN2}J.~Marsano, N.~Saulina and S.~Sch\"afer-Nameki, ``G-flux, M5 instantons, and U(1) symmetries in F-theory'', {\it Phys. Rev.} {\bf D87} (2013) 066007 [arXiv:1107.1718 [hep-th]].
\bibitem{BCV}A.~P.~Braun, A.~Collinucci and R.~Valandro, ``G-flux in F-theory and algebraic cycles'', {\it Nucl. Phys.} {\bf B 856} (2012) 129 [arXiv:1107.5337 [hep-th]].
\bibitem{MS}J.~Marsano and S.~Sch\"afer-Nameki, ``Yukawas, G-flux, and Spectral Covers from Resolved Calabi-Yau's'', {\it JHEP} {\bf 11} (2011) 098 [arXiv:1108.1794 [hep-th]].  
\bibitem{KMW}S.~Krause, C.~Mayrhofer and T.~Weigand, ``$G_4$ flux, chiral matter and singularity resolution in F-theory compactifications'', {\it Nucl. Phys.} {\bf B 858} (2012) 1--47 [arXiv:1109.3454 [hep-th]].
\bibitem{GH}T.~W.~Grimm and H.~Hayashi, ``F-theory fluxes, Chirality and Chern-Simons theories'', {\it JHEP} {\bf 03} (2012) 027 [arXiv:1111.1232 [hep-th]].
\bibitem{KMW2}S.~Krause, C.~Mayrhofer and T.~Weigand, ``Gauge Fluxes in F-theory and Type IIB Orientifolds'', {\it JHEP} {\bf 08} (2012) 119 [arXiv:1202.3138 [hep-th]]. 
\bibitem{IJMMP}K.~Intriligator, H.~Jockers, P.~Mayr, D.~R.~Morrison and M.~R.~Plesser, ``Conifold Transitions in M-theory on Calabi-Yau Fourfolds with Background Fluxes'', {\it Adv. Theor. Math. Phys.} {\bf 17} (2013) 601 [arXiv:1203.6662 [hep-th]].
\bibitem{KSN}M.~Kuntzler and S.~Sch\"afer-Nameki, ``	
G-flux and Spectral Divisors'', {\it JHEP} {\bf 11} (2012) 025 [arXiv:1205.5688 [hep-th]].
\bibitem{CGK}M.~Cveti\v c, T.~W.~Grimm and D.~Klevers, ``Anomaly Cancellation And Abelian Gauge Symmetries In F-theory'', {\it JHEP} {\bf 02} (2013) 101 [arXiv:1210.6034 [hep-th]]. 
\bibitem{BCV2}A.~P.~Braun, A.~Collinucci and R.~Valandro, ``Hypercharge flux in F-theory and the stable Sen limit'', {\it JHEP} {\bf 07} (2014) 121 [arXiv:1402.4096 [hep-th]].
\bibitem{SNW}S.~Sch\"afer-Nameki and T.~Weigand, ``F-theory and 2d (0,2) Theories'', {\it JHEP} {\bf 05} (2016) 059 [arXiv:1601.02015 [hep-th]].

\bibitem{Nak}N.~Nakayama, ``On Weierstrass Models'', {\it Algebraic Geometry and Commutative Algebra in Honor of Masayoshi Nagata}, (1988), 405--431.  
\bibitem{DG}I.~Dolgachev and M.~Gross, ``Elliptic Three-folds I: Ogg-Shafarevich Theory'', {\it Journal of Algebraic Geometry} {\bf 3}, (1994), 39--80.
\bibitem{G}M.~Gross, ``Elliptic Three-folds II: Multiple Fibres'', {\it Trans. Amer. Math. Soc.} {\bf 349}, (1997), 3409--3468.

\bibitem{DWmodel}R.~Donagi and M.~Wijnholt, ``Model Building with F-Theory'', {\it Adv. Theor. Math. Phys.} {\bf 15} (2011) no.5, 1237--1317 [arXiv:0802.2969 [hep-th]].
\bibitem{BHV1}C.~Beasley, J.~J.~Heckman and C.~Vafa, ``GUTs and Exceptional Branes in  F-theory -I'', {\it JHEP} {\bf 01} (2009) 058 [arXiv:0802.3391 [hep-th]].
\bibitem{BHV2}C.~Beasley, J.~J.~Heckman and C.~Vafa, ``GUTs and Exceptional Branes in F-theory - II: Experimental Predictions'', {\it JHEP} {\bf 01} (2009) 059 [arXiv:0806.0102 [hep-th]].
\bibitem{DW}R.~Donagi and M.~Wijnholt, ``Breaking GUT Groups in F-Theory'', {\it Adv. Theor. Math. Phys.} {\bf 15} (2011) 1523--1603 [arXiv:0808.2223 [hep-th]].

\bibitem{Cas}J.~W.~S.~Cassels, {\it Lectures on Elliptic Curves}, London Math. Society Student Texts {\bf 24}, Cambridge University Press (1991).

\bibitem{BIKMSV}M.~Bershadsky, K.~A.~Intriligator, S.~Kachru, D.~R.~Morrison, V.~Sadov and C.~Vafa, ``Geometric singularities and enhanced gauge symmetries'', {\it Nucl. Phys.} {\bf B 481} (1996) 215 [arXiv:hep-th/9605200].

\bibitem{GKK}T.~W.~Grimm, A.~Kapfer and D.~Klevers, ``The Arithmetic of Elliptic Fibrations in Gauge Theories on a Circle'', {\it JHEP} {\bf 06} (2016) 112 [arXiv:1510.04281 [hep-th]].


\end{thebibliography}
\end{document}